# On the miscibility gap at 800 K in the Fe-Cr alloy system

S. M. Dubiel[1*] and J. Żukrowski[2]

[1]AGH University of Science and Technology, Faculty of Physics and Applied Computer Science, al. A. Mickiewicza 30, 30-059 Kraków, Poland, [2]AGH University of Science and Technology, Academic Center for Materials and Nanotechnology, al. A. Mickiewicza 30, 30-059 Kraków, Poland

## Abstract

Issues pertinent to the miscibility gap at 800K in a $Fe_{73.7}Cr_{26.3}$ alloy viz. the kinetics of the phase decomposition and borders of the latter were investigated by means of the Mössbauer-effect spectroscopy. The kinetics was revealed to well follow the Johnson-Mehl-Avrami-Kolgomarov equation testifying to a nucleation and growth mechanism underlying the decomposition. A solubility limit of Cr in the Fe matrix was determined to be 17.3(6) at.% and that of Fe in the Cr matrix was found as equal to 18.5(9) at.%. The results obtained have been compared with various theoretical predictions.

* Corresponding author: **Stanislaw.Dubiel@fis.agh.edu.pl**

Fe-Cr alloys have been of industrial and scientific interests since over a century. The former stems from their importance in the steel making industry while the latter from their interesting magnetic properties. Consequently they have been regarded as model alloys and were subject of numerous experimental and theoretical studies. An extensive review on the Fe-Cr alloy system was recently given by Xiong et. al. [1]. Despite all these studies a crystallographic phase diagram of the system continues to be a subject of both experimental and theoretical studies as borders of a miscibility gap and those of a sigma phase are not precisely known and substantial differences exist between different theoretical treatments [2-6]. The issue is not only of interest *per se* but also because of technological importance of steels produced on the basis of the Fe-Cr alloys. Due to their superior properties, like resistance to high-temperature corrosion, various devices fabricated therefrom work at service at elevated temperatures at which they deteriorate. Two phenomena related with the crystallographic phase diagram of Fe-Cr are responsible for these materials worsening viz. (1) phase decomposition into Fe-rich ($\alpha$) and Cr-rich ($\alpha'$) phases, and (2) sigma-phase precipitation. The phase decomposition occurs within the so-called miscibility gap (MG) and its metallurgical consequences are known as the 475$^{o}$C embrittlement. Of special practical interest is the Fe-rich border of the miscibility gap i.e. the solubility limit of Cr in Fe matrix as its value is close to a minimum Cr content above which the Fe-Cr alloy becomes stainless. The present study was aimed at determining both Fe-rich and Cr-rich borders at 800 K i.e. the temperature at which there are meaningful differences between various theoretical predictions [2-6]. Mössbauer spectroscopy was successfully applied as relevant tool for determining borders of the MG [7-10].

The investigation was carried out on a ~25 µm thick $Fe_{73.7}Cr_{26.3}$ alloy foil in form a 20x20 mm rectangle obtained by a cold rolling a 100 µm thick tape. Details of preparation of the latter are given elsewhere [11]. $^{57}$Fe Mössbauer spectra, of which some examples are shown in Fig. 1, were measured at room temperature in transmission geometry, using a standard spectrometer with a drive working in a sinusoidal mode. 14.4 keV gamma rays were supplied by a $^{57}$Co/Rh source. Each spectrum was registered in 1024 channels of a multichannel analyzer within a 2 days run.

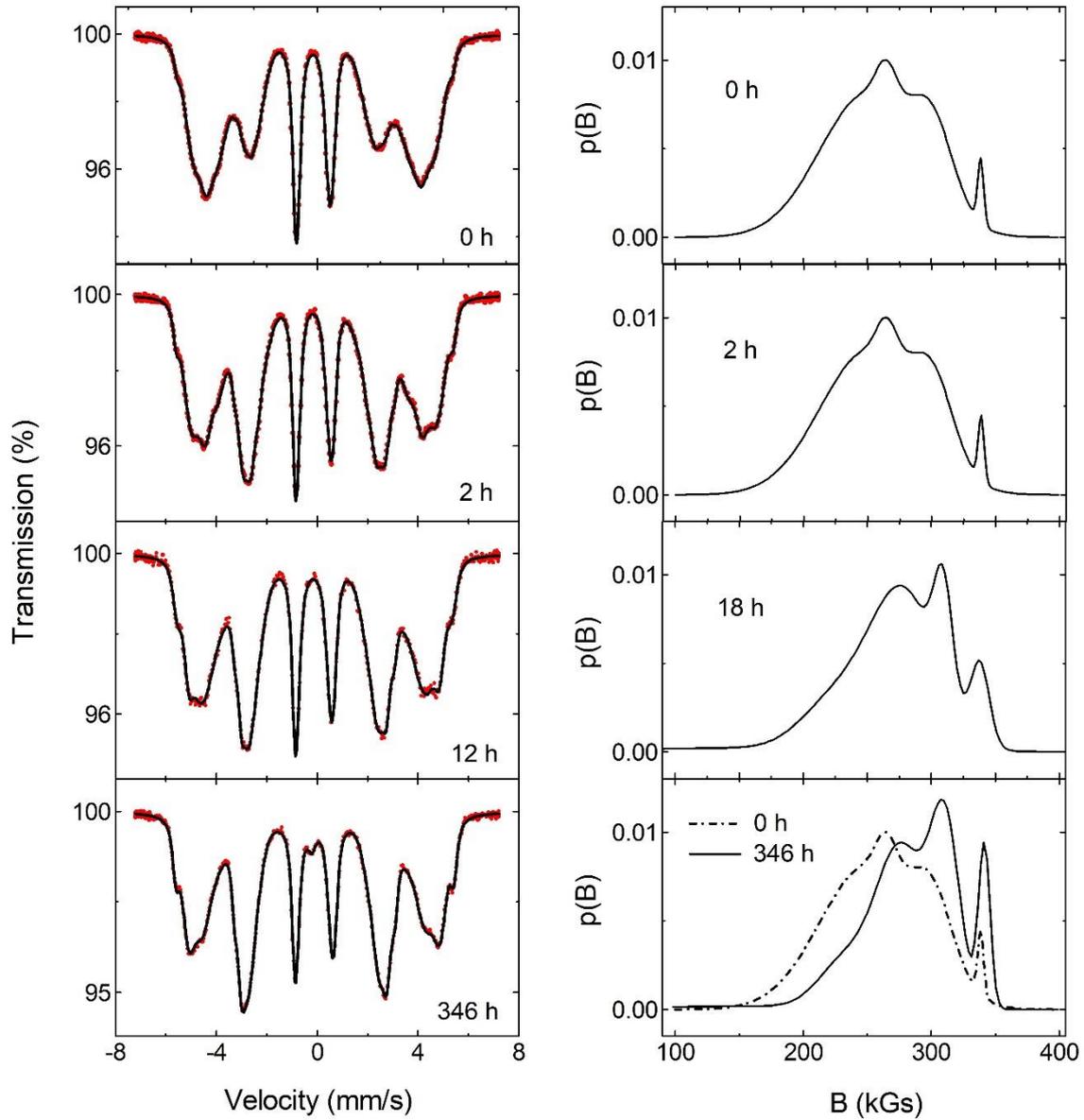

Fig. 1. (left panel) Examples of $^{57}$Fe Mössbauer spectra recorded at RT on a Fe$_{73.7}$Cr$_{26.3}$ sample annealed in vacuum at 800K for various periods indicated. Note a single-line contribution in the central part of the bottom-most spectrum. It is associated with the Cr-rich α' phase. Corresponding distribution curves of the hyperfine field derived from the spectra are shown in the right panel where on the bottom plot a comparison between the extreme cases is presented.

The spectra were analyzed assuming that an effect of presence of Cr atoms in the first-two neighbor shells around $^{57}$Fe probe nuclei, *1NN-2NN*, on the hyperfine field, *B,* and on the isomer shift, *IS*, was additive i.e. $X(m,n) = X(0,0) + m\Delta X_1 + n\Delta X_2$, where *X=B or IS, ΔX$_k$* is a change of *X* due to one Cr atom situated in *1NN* (*k*=1) or in *2NN* (*k*=2). This procedure has proved to work properly in different binary alloys of Fe including Fe-Cr ones e. g. [8-11]. The total number of possible atomic configurations *(m,n)* in the two-shell approximation is equal to 63, but for *x* = 26.3 at% most of them have vanishingly small probabilities, so 17 most probable (according to the binomial distribution) were chosen to be included into the fitting procedure (their overall probability was > 0.97). However, their probabilities (associated with spectral areas of sextets corresponding to the chosen configurations) were treated in the fitting procedure as free parameters. Free parameters were also *X(0,0),* line width, *G*, $\Delta B_1$, $\Delta B_2$, and relative ratio of the lines within the sextets. On the other hand, fixed were values of $\Delta IS_1$ = -0.02 mm/s, and $\Delta IS_1$= -0.01 mm/s. In the case of the spectra annealed for the three longest periods viz. 83, 203 and 346.5 hours additional singlet was included to account for a single-line sub spectrum observed in the central part of the corresponding spectra. This sub spectrum has been associated with the Cr-rich phase ($\alpha'$) precipitated as a result of the phase decomposition. In addition, some spectra were analyzed in terms of a hyperfine distribution method [12]. The right-hand panel in Fig. 1 shows examples of the hyperfine distribution curves delivered by the latter procedure. They nicely illustrate a redistribution of Cr atoms that had taken place in the course of the applied annealing, and, in particular, a shift of the distribution towards a higher value is evident testifying to a clustering of Cr atoms.

A selection of the best-fit spectral parameters obtained with the two-shell model is displayed in Table 1. Two quantities can be readily used to quantitatively follow a kinetics of the phase decomposition viz. the average hyperfine field, $<B> = \sum_{m,n} P(m,n) \cdot B(m,n) / \sum_{m,n} P(m,n)$, and the *P(0,0)* atomic configuration i.e. the one without Cr atoms within the *1NN-2NN* vicinity of the probe Fe atoms. A dependence of the former on the annealing time can be seen in Fig. 2.

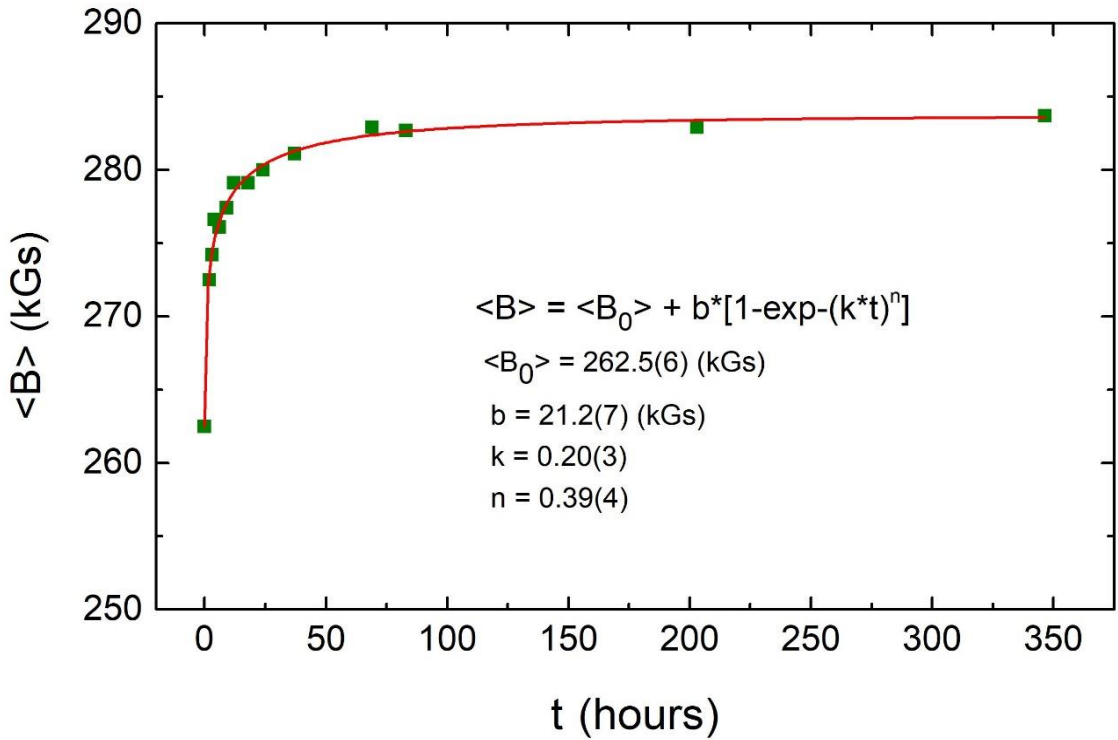

Fig. 2 The average hf. field, *<B>*, versus the annealing time, *t*. The best-fit of the JMAK equation to the data is shown as full line, and the best-fit parameters are indicated, too. Avrami exponent is depicted by *n* and the rate constant by *k*.

The measured *<B>*-values are in line with the Johnson-Mehl-Avrami-Kolgomarov equation (JMAK) equation testifying thereby to a nucleation and growth origin of the decomposition process. The value of the Avrami exponent, *n*, is close to 0.5, indicating by that the process responsible for the phase decomposition in this sample could be a diffusion-controlled thickening of plates after their edges have impinged [13]. Noteworthy, similar *n*-values were found for a $Fe_{85}Cr_{15}$ sample annealed at 688 and 723 K [9].

Interestingly, the annealing time evolution of the *P(0,0)*-values shows a similar behavior as the *<B>(t)* one – see Fig. 3. Consequently, it was also fitted to the JMAK-like equation, and the kinetics parameters have, within error limits, the same values as those determined from the *<B>(t)*-dependence.

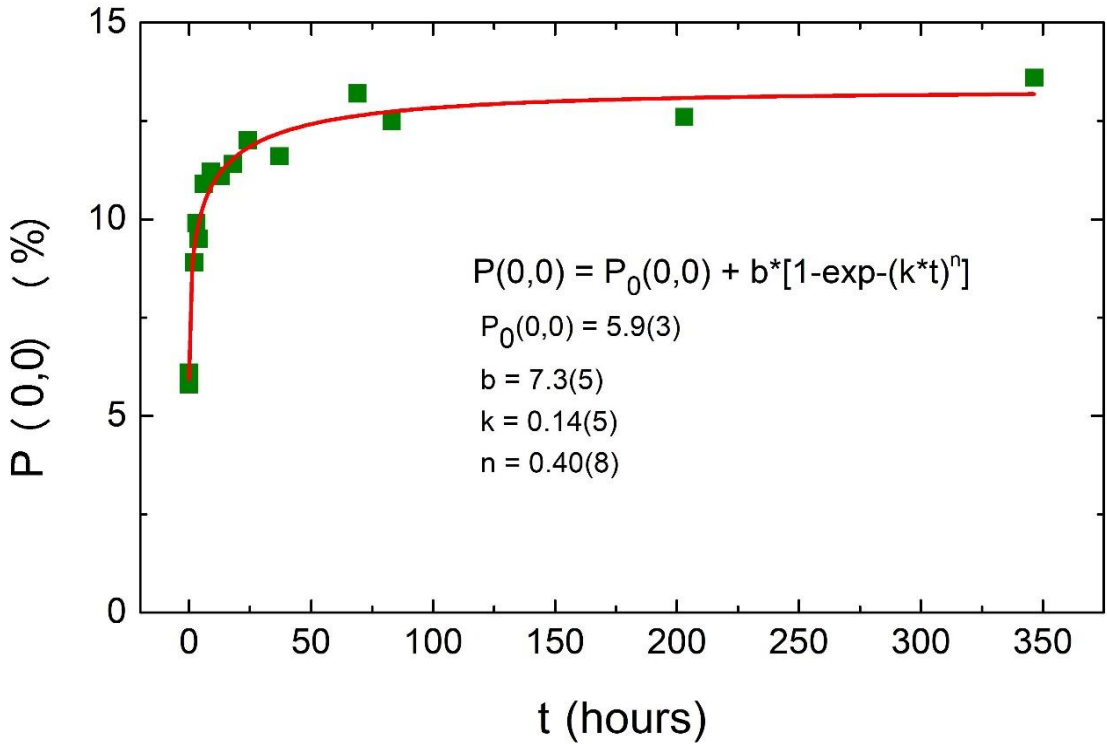

Fig. 3 The *P(0,0)* versus the annealing time, *t*. The best-fit of the JMAK equation to the data is presented as full line, and the best-fit parameters are indicated, too.

The *P(0,0)*-values can be used to estimate Cr content in the Fe matrix. For a random distribution the following relationship holds:

$$x = 1 - \sqrt[14]{P(0,0)} \qquad (1)$$

For the untreated alloy $P(0,0)=0.059$ which yields $x=18.3(3)$ at% Cr via equ. (1). This is significantly less than the Cr concentration obtained from the chemical analysis viz. 26.3 at%. However, the latter is an average Cr concentration in the bulk, while the former gives its local value as seen by one particular atomic configuration. In any case, the discrepancy means that the distribution of Cr atoms in the initial state of the studied alloy was not random. In the course of annealing i.e. the phase decomposition the *P(0,0)*-values grow reaching the value of 0.13 (an average over the last three annealing times). The latter figure via the equ. (1) yields $x=13.5(4)$ at.% Cr. That means that the Cr concentration, as seen by the Fe atoms having no Cr atoms within their 2-shells neighborhood, decreased by 4.8(7) at%. This effect can be understood in terms of Cr atoms clustering i.e. formation of the Cr-rich α′ phase. A confirmation of this effect follows also from a decrease of the average number of Cr atoms within the *1NN-2NN* vicinity around the Fe probe atoms, <m+n>, on the annealing time – see Table 1.

The <B>-value in saturation equals to 283.7 kGs can be readily used to determine a solubility limit of Cr in the Fe matrix, $x_{Cr}$. For that purpose one uses a monotonous dependence of <B> on the composition in Fe-rich Fe-Cr alloys [8], and arrives at $x_{Cr}$=17.3(6) at.%. On the other hand, the value of the isomer shift, IS=-0.21 mm/s (relative to a Co/Rh source) or -0.09 mm/s (relative to a metallic Fe) can be used to determine a concentration of Fe in the Cr-rich α' phase, $x_{Fe}$. To that end one applies a relationship between the isomer shift and composition as found for Cr-rich Fe-Cr alloys [8] arriving at $x_{Fe}$=18.5 (9) at.%. These values can be next compared with theoretical predictions relevant to the miscibility gap (MG) in the Fe-Cr alloy system in order to validate them. Thus in Fig. 4 such comparison is made with the predictions by Bonny et al. [5]. Large errors of the potential method do not allow distinguishing between the two predictions. Measurements at temperatures > 800 K are needed.

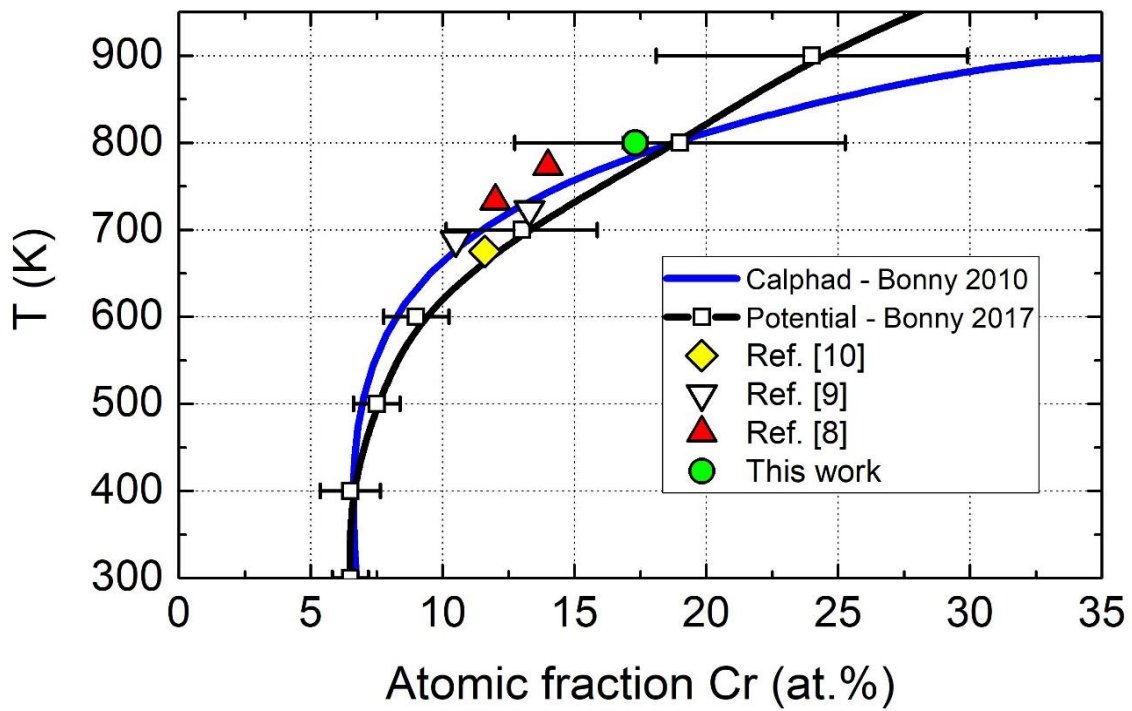

Fig. 4 A comparison between two theoretical predictions concerning the Fe-rich branch of the MG [5] and experimental data obtained with the Mössbauer spectroscopy.

Further comparison of our presently and earlier obtained experimental data can be made with calphad-based calculations of the full MG [2,4,6]. The corresponding juxtaposition is displayed in Fig. 5. Unfortunately, the data cover the temperature range where differences between the predictions are rather minor, making their validation not unique. The largest differences between the predictions exist at T < 650 K and at T > 800 K, hence the future validation of them should be done either in the former or in the latter temperature range. However, the rate of the phase decomposition dramatically decreases with T [10], which means that in practice only measurements at T > 800 K are feasible. Nevertheless, it can be already stated that the presently obtained data for the Fe-rich branch of the MG does not agree with the prediction made by Andersson and Sundman [2] according to which the Cr solubility limit in Fe at 800 K is equal to 21 at.% which is significantly more than the value of 17.3(6) at.% determined in this study. On the other hand, the presently found value of the Fe solubility limit in Cr at 800 K agrees perfectly with the value calculated by these authors [2].

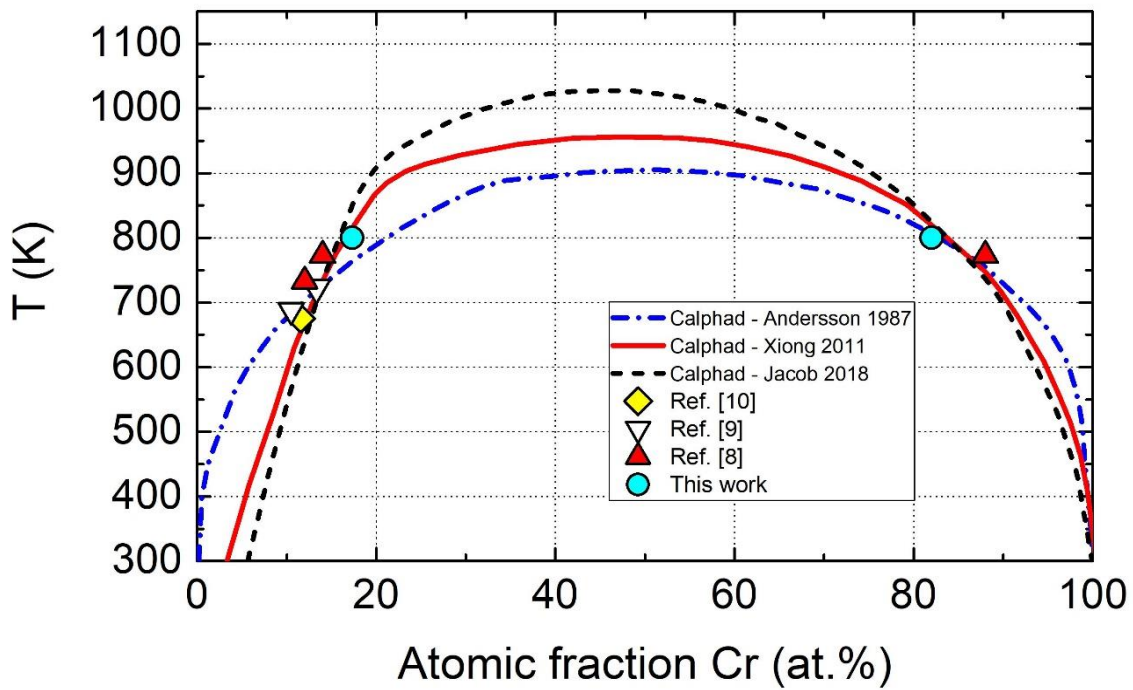

Fig. 5 A comparison between three theoretical calphad-based predictions concerning the full MG and our experimental data obtained with the Mössbauer spectroscopy.

Finally, the analysis of the spectra yielded also information on the average angle between the direction of a magnetic moment and that of the gamma-rays (normal to the sample's surface), $\Theta$. As shown in Fig. 6, $\Theta$ increases by ~19° within the 346 hours of annealing, yet after the first two hours of the annealing $\Theta$ rotated from 45° to 58° to gradually increase in the course of subsequent annealing time. The initial jump likely reflects a release of strain and the following slow increase of $\Theta$ is related to a recrystallization process. Noteworthy, the data are in line with the JMAK equation.

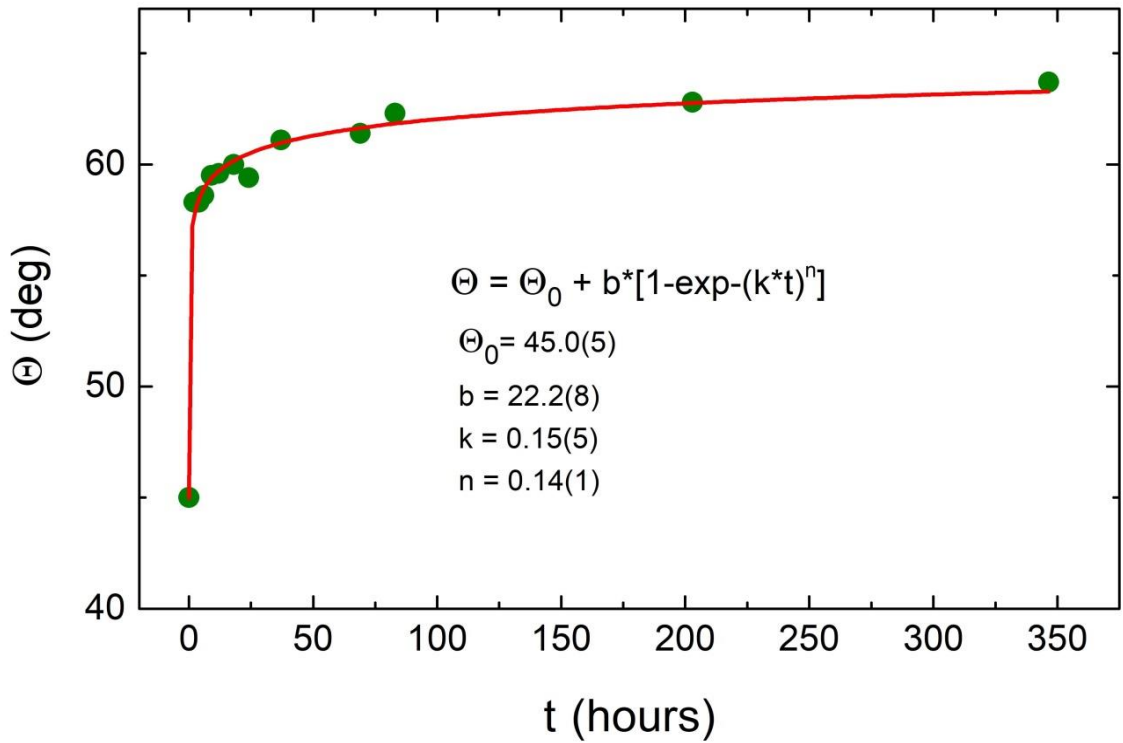

Fig. 6 An average angle, $\Theta$, between the sample magnetic moment and the gamma rays direction (normal to the sample's surface), vs. annealing time, $t$.

In summary, the kinetics of the phase decomposition at 800 K in a $Fe_{73.7}Cr_{26.3}$ alloy was determined and successfully described in terms of the JMAK equation. Borders of the miscibility gap have been determined and compared with various theoretical predictions. Measurements at T > 800 K are needed to validate these predictions more thoroughly.

**Acknowledgements**

This work was financed by the Faculty of Physics and Applied Computer Science AGH UST and ACMIN AGH UST statutory tasks within subsidy of Ministry of Science and Higher Education. G. Bonny is gratefully thanked for making available his calculations shown in Fig. 4.

Table 1 Best-fit spectral parameters as obtained for the studied $Fe_{73.7}Cr_{26.3}$ alloy. The meaning of the parameters is given in the text. The isomer shift values are given relative to the Co/Rh source.

| t [h] | P(0,0) [%] | B(0,0) [kGs] | $\Delta B_1$ [kGs] | $\Delta B_2$ [kGs] | <B> [kGs] | <m+n> | <IS> [mm/s] | G [mm/s] | $\Theta$ [°] | $IS_{\alpha'}$ [mm/s] | $G_{\alpha'}$ [mm/s] | $A_{\alpha'}$ [%] |
|---|---|---|---|---|---|---|---|---|---|---|---|---|
| 0 | 5.8(4) | 332.9(8) | -33.5 | -21.4 | 261.8 | 2.54 | -0.136 | 0.35(4) | 45.0 | | | |
| 2 | 8.9(3) | 336.6(3) | -33.1 | -21.7 | 271.5 | 2.35 | -0.138 | 0.33(3) | 58.3 | | | |
| 3 | 9.9(3) | 336.7(3) | -33.6 | -22.1 | 272.9 | 2.33 | -0.134 | 0.32(3) | 58.3 | | | |
| 4 | 9.5(4) | 336.1(4) | -31.8 | -21.3 | 274.4 | 2.34 | -0.134 | 0.32(3) | 58.3 | | | |
| 6 | 10.9(5) | 336.7(4) | -32.5 | -21.4 | 275.2 | 2.25 | -0.136 | 0.34(3) | 58.6 | | | |
| 9 | 11.2(5) | 336.2(4) | -32.8 | -21.6 | 276.5 | 2.16 | -0.136 | 0.33(3) | 59.5 | | | |
| 13 | 11.1(5) | 336.5(4) | -32.2 | -21.4 | 277.4 | 2.11 | -0.135 | 0.33(3) | 59.6 | | | |
| 18 | 11.4(4) | 337.3(3) | -32.3 | -21.2 | 278.2 | 2.20 | -0.133 | 0.33(3) | 60.0 | | | |
| 24 | 12.0(3) | 338.0(2) | -33.0 | -21.9 | 280.0 | 2.02 | -0.134 | 0.33(3) | 59.4 | | | |
| 37 | 11.6(5) | 337.8(3) | -31.2 | -20.4 | 281.1 | 2.16 | -0.132 | 0.33(3) | 61.1 | | | |
| 69 | 13.2(5) | 338.0(4) | -30.9 | -20.6 | 282.9 | 2.15 | -0.134 | 0.33(3) | 61.4 | | | |
| 83 | 12.5(5) | 339.8(5) | -30.0 | -19.8 | 283.5 | 1.96 | -0.132 | 0.32(4) | 62.3 | -0.21(2) | 0.31(5) | 0.8(1) |
| 203 | 12.6(4) | 339.4(3) | -30.1 | -20.0 | 282.9 | 2.14 | -0.132 | 0.34(3) | 62.8 | -0.22(2) | 0.31(5) | 1.0(1) |
| 346.5 | 13.6(3) | 339.1(4) | -31.4 | -20.7 | 283.7 | 2.01 | -0.132 | 0.33(5) | 63.7 | -0.21(1) | 0.30(5) | 1.1(1) |